\documentstyle[12pt,epsf]{article}
\newcommand{\be}{\begin{equation}}
\newcommand{\ee}{\end{equation}}
\newcommand{\bea}{\begin{eqnarray}}
\newcommand{\beas}{\begin{eqnarray*}}
\newcommand{\eea}{\end{eqnarray}}
\newcommand{\eeas}{\end{eqnarray*}}
\newcommand{\ba}{\begin{array}}
\newcommand{\ea}{\end{array}}
\newcommand{\ov}{\overline}
\newcommand{\eq}{\setcounter{equation}{0}}
\newcommand{\la}{\langle}
\newcommand{\ra}{\rangle}

\newcommand{\rar}{\rightarrow}

\newcommand{\Rar}{\longrightarrow}
\newcommand{\com}{ C\hspace{-1.5ex}l~}

\def\etal{{\it et. al.}}
\def\pl{{\it Phys.\ Lett.\ }}

\def\prd{{\it Phys.\ Rev.\ D }}

\def\prl{{\it Phys.\ Rev.\ Lett.\ }}

\def\thebibliography#1{\section*{\refname}\list
 {\arabic{enumi}. }{\settowidth\labelwidth{[#1]}\leftmargin\labelwidth
 \advance\leftmargin\labelsep
 \usecounter{enumi}}
 \def\newblock{\hskip .11em plus .33em minus .07em}
 \sloppy\clubpenalty4000\widowpenalty4000
 \sfcode`\.=1000\relax}

\begin{document}

\title{Topological Defects in $[SU(6)]^3 \times Z_3$}

\author{ H. Garc\'{\i}a-Compe\'an, A. P\'erez-Lorenzana and
A. Zepeda\\[2ex]
{\it Departamento de F\'{\i}sica}\\
{\it Centro de Investigaci\'on y de Estudios Avanzados del I.P.N.}\\
{\it Apdo. Postal 14-740, 07000, M\'exico, D. F., M\'exico.}}

\date{ }

\maketitle

\begin{abstract}
We study topological defects arising in the Grand Unification Mo\-del
$SU(6)_L\otimes SU(6)_c \otimes SU(6)_R \times Z_3$. We show that the
model does not contain domain walls, while it produces
massive magnetic monopoles and it may, depending on the symmetry breaking chain,
give rise to the formation of strings.  We also discuss their possible relation with
the origin of the highest energy cosmic rays detected.\\[1ex]
\end{abstract}

\newpage

\section{Introduction}

The sources of the  high energy cosmic ray (HECR) events recently  observed
above $10^{20}\, eV$ ($100\ EeV$) by the Fly's Eye~\cite{fly},
AGASA~\cite{agasa}, Haverah Park~\cite{haverah} and Yakutsk~\cite{yakutsk}
experiments remain unknown. Conventional astrophysical accelerator mechanisms
encounter severe difficulties in accelerating particles to these
energies~\cite{accel}. It is hard to accelerate protons and heavy nuclei up to
such energies even in the most powerful astrophysical objects. Also, the GZK
cut-off~\cite{gzk} limits the distance to a possible source of nucleons with
energy above $\simeq 70\, EeV$ to less than $\simeq 100 Mpc$~\cite{sigl1} from
the Earth. Other possible primary candidates for these HECR events could be
gamma rays and neutrinos. Nevertheless, the gamma-ray hypothesis appears
inconsistent with the temporal distribution of the Fly's Eye
event~\cite{halzen}, and also the density profile of the Yakutsk event showed a
large number of muons, which argues against this hypothesis~\cite{yakutsk}.
Moreover the mean free path for a $10^{20} eV$ photon before it annihilates on
the microwave background into a $e^-e^+$ pair  is around  10 to $40Mpc$. On the
other hand, the Fly's eye event occurred high in the atmosphere, whereas the
expected event rate for early development of a neutrino induced air shower is
down from that of an electromagnetic or hadronic interaction by six orders of
magnitude~\cite{halzen}.

These difficulties have motivated two recent suggestions. The first is  that
the underlying production mechanism of HECR could be of  a non acceleration
nature, namely the decay of supermassive elementary particles related to Grand
Unified Theories (GUT's). Such particles could be released from topological
defects (TD), such as cosmic strings, monopoles and  domain walls, which could
be formed during the phase transitions associated with the spontaneous symmetry
breaking (SB)  of the GUT symmetry in the early
universe~\cite{sigl1,td1,kibble}. TD are topologically stable but nevertheless
they can release part of their energy in the form of supermassive elementary
particles (which could subsequently decay into leptons and quarks) due to
physical processes like collapse or annihilation. TD are therefore  viable
sources of  HECR, they predict injection spectra which are considerably harder
than shock acceleration spectra and also there is no cut-off effect in the
attenuation of ultra high energy $\gamma$-rays which dominate the predicted
flux~\cite{sigl1,sigl2,chi}, although the absolute flux levels predicted by TD
are model dependent~\cite{gill}.  The other  suggestion is  that the
primary particles of the HECR may be relativistic magnetic monopoles, with
masses bounded by $M \leq 10^{10\pm 1}\ GeV$, to be consistent with the Parker
limit and other phenomenological bounds~\cite{kephart,parker}. Both suggestions
deal with some kind of TD. But in general TD models are very constrained by
their astrophysical implications.  Supermassive monopoles are perceived to be
enough of a cosmological problem~\cite{preskill,tdvil}, so that they have to be
``inflated'' away~\cite{guth}. Also they can catalyze  proton
decay~\cite{rubakov}. Conventional GUT's, where the electroweak and strong
forces are unified into a symmetry group, which is broken down to the standard
one at an energy scale about $10^{16}~GeV$~\cite{guts}, imply an over abundance
of supermassive monopoles formed through the Kibble mechanism~\cite{kibble}. 
Nevertheless, if one assumes that monopoles exist at the level of abundance
compatible with known experimental~\cite{ahlen} and phenomenological upper
bo\-unds~\cite{parker}, they could give an important contribution to the HECR
flux~\cite{sigl1,sigl2}.

Domain walls are believed to be cosmological disasters and a particle physics
model is considered inadmissible if it predicts them~\cite{domain}.
Superconducting cosmic strings~\cite{witten} can not produce the HECR flux at
the present epoch without violating the $^4$He--photodisintegration
bound~\cite{he4}. Cosmic strings, on the other hand, are considered 
astrophysically promising~\cite{cite}, and a viable source of HECR.

In this way, TD arising from an specific Grand Unification Model (GUM) come to
be of interest. Moreover, the restrictions on TD and their astrophysical
implications form a phenomenological test for all the GUM's proposed until
now.  Some work on this issue have been done in the most popular models in the
past (see for example references ~\cite{so10}). This leads us to study  the
possible contributions to HECR from  TD arising in a new GUT model proposed in
recent years~\cite{oax,2,tuning}, which is based on the gauged symmetry
$[SU(6)]^3\times Z_3$ and which has a set of properties which make it  a viable
proposal for the symmetry of the non-gravitational interactions in  nature. In
the present article we develop the first part of this program; here we 
analyze the SB patterns in the model and establish what kind of TD are
involved.

In section 2 we review the $[SU(6)]^3
\times Z_3$ Grand Unification Model following basically  Refs.
\cite{oax,2,tuning} paying attention to its properties and advantages.
In section 3 we discuss the absence of domain walls and cosmic strings
in the model. This occurs even if the SB pattern keeps the discrete
symmetry $Z_3$
intact in the first step.  However as we will see, cosmic strings can
arise in the model if different symmetry breaking chains are considered.
Thus for the symmetry breaking involved the only TD generated
in the phase
transitions of the model are monopoles and textures. This is discussed
in section 4, where we  also  
discuss briefly the possible relation of monopoles with the mechanisms of
production of
HECR mentioned in the second paragraph.
Finally in section 5 we give some concluding remarks. One
appendix at the end of the paper deals with details  of the SB
implemented in section 3.

\section{Brief review of $\bf [SU(6)]^3\times Z_3$}
\eq

The model under consideration is based on the gauge group \be  G \equiv
SU(6)_L\otimes SU(6)_c\otimes SU(6)_R\times Z_3 \ee and unifies
non-gravitational forces with transitions among three families.  In Eq. (2.1)
$\otimes$ indicates a direct product, $\times$ a semidirect one, and $Z_3$ is a
three-element cyclic group acting upon $[SU(6)]^3$ such that if $(A,B,C)$ is a
representation of $[SU(6)]^3$ with $A$ a representation of the first factor,
$B$ of the second and $C$ of the third, then $Z_3(A,B,C)\equiv (A,B,C) \oplus
(B,C,A) \oplus (C,A,B)$ is a representation of $G$. $SU(6)_c$ is a vector-like
group which includes three quark-like colors  and three lepton-like ones, and
it includes as a subgroup the $SU(3)_c \otimes U(1)_{B-L}$ group of the
left-right symmetric (LRS) extension of the Standard Model (SM).
$SU(6)_L\otimes SU(6)_R$ includes the $SU(2)_L\otimes SU(2)_R$ gauge group of
the LRS model.

Among the special properties of this model we may recall that its gauge group,
$G$, is the maximal unifying group for the three families, with left-right
symmetry and with (extended) vector color and that it leads to absolute
(perturbative) stability of the proton~\cite{mass}. The quark-lepton symmetry
in this model is maximal, since it contains as many leptons colors as quarks
colors. Furthermore, all the  fermions in the model, including the known ones,
belong to a single irreducible representation (irrep) of $G$. On the other hand
the presence of the horizontal group in $SU(6)_L\otimes SU(6)_R$ allows the
possibility of obtaining predictions for the fermion mass matrices
\cite{2,tuning}. Besides its aesthetic appeal, the viability of the model
steams from its capacity to match the observed values of the SM couplings
constants (see below). For these reasons we consider it of interest to analyze
the TD properties of its SB chain.

The 105 gauge fields (GF's) in $G$ can be divided in two sets: 70 of them
belonging to $SU(6)_L\otimes SU(6)_R$ and 35 being associated with $SU(6)_c$.
The first set includes $W^\pm_L$ and $W^0_L$ (the GF's of the known weak
interactions), the GF's associated with $SU(2)_R$, the GF's of the horizontal
interactions, and the GF's of the nonuniversal charged and neutral
interactions. All of them have electrical charges $0$ or $\pm 1$. The
generators of $SU(6)_{L(R)}$ may be written in a
$SU(2)_{L(R)}\otimes SU(3)_{HL(HR)}$ basis as 
\be \sigma_i\otimes I_3/2\sqrt{3},\qquad I_2\otimes\lambda_\alpha/2\sqrt{2},
\qquad \sigma_i\otimes\lambda_\alpha/2\sqrt{2},    \label{LRgenerators} 
\ee
where $\sigma_i$ are the $2\times 2$ Pauli matrices, $\lambda_\alpha$ are the
$3\times 3$ Gell-Mann matrices, and $I_2$ and $I_3$ are the $2\times 2$ and
$3\times 3$ identity matrices respectively. The second set includes the eight
gluon fields of $SU(3)_c$, nine lepto-quarks ($X_i$, $Y_i$ and $Z_i$,
$i=1,2,3$, with electrical charges $-2/3$, $1/3$ and $-2/3$ respectively),
their nine conjugated, six dileptons ($P^\pm_a$, $P^0$ and $\tilde P^0$,
$a=1,2$, with electrical charges as indicated), plus the GF's associated with
diagonal generators in $SU(6)_c$ which are  not taken into account already in
$SU(3)_c$.

The fermions of the model are in the irrep 108,
\be
\psi(108)_L = Z_3\psi (6,1,\ov 6)_L =\psi(6,1,\ov 6)_L \oplus
\psi(\ov 6,6,1)_L \oplus \psi(1,\ov 6,6)_L,
\ee
with quantum numbers with respect to $(SU(3)_c, SU(2)_L, U(1)_Y)$ given by
\beas
\psi(\ov 6,6,1)_L &\equiv& \psi^\alpha_a:  \ \ \ \ \ 3(3,2,1/3) \oplus
6(1,2,-1) \oplus 3(1,2,1),\\
\psi(1,\ov 6,6)_L &\equiv& \psi^A_\alpha:  \ \ \ \ \ 3(\ov 3,1,-4/3) \oplus
3(\ov 3,1,2/3)\oplus 6(1,1,2) \oplus 9(1,1,0)\\
& &  \ \ \ \qquad \oplus 3(1,1,-2),\\
\psi(6,1,\ov 6)_L & \equiv& \psi^a_A:  \ \ \ \ \ 9(1,2,1) \oplus 9(1,2,-1),
\eeas
where $a,b,\dots,A,B,\dots,\alpha,\beta,\dots=1,\dots, 6$ label $L$,
$R$ and $C$ tensor indices, respectively. The known fermions are
contained in the $\psi(\ov 6,6,1)_L\oplus\psi(1,\ov{6},6)_L$ part of
$\psi(108)$.

In order to achieve the SB we introduce
appropriate Higgs scalars. Using the branching rules
\beas
SU(6)_{L(R)}& \rightarrow &
SU(2)_{L(R)}\otimes SU(3)_{HL(R)}\\
 6 & \rightarrow & (2,3)\\
 15 & \rightarrow & (1,6) \oplus (3, \ov{3} )\\
 21 & \rightarrow & (1,\ov{3}) \oplus (3,6)
\eeas
 and
\beas
SU(6)_c& \rightarrow & SU(3)_c\\
 6  & \rightarrow & (3) + 3 (1)\\
 15 & \rightarrow & (\ov {3}) + 3(3)+ 3(1) \\
 21 & \rightarrow & (6) + 6(1) + 3(3),
\eeas
we can see that the vacuum expectation values (vevs) of a 6 of $SU(6)_L$
necessarily break $SU(2)_L$.  However, the $Z_3$ symmetrized version of a 6,
$\phi(18) = Z_3 \phi(6,1,1)$, is not sufficient to give tree-level mass to at
least one ordinary fermion. We therefore assume that the last step in the SB
chain of $G$ is due to the vevs of a $\phi_4 = \phi(108) = Z_3
\phi(1,\ov{6},6)$ and that these vevs lie only in the electrically neutral
directions in the $SU(6)_L \otimes SU(6)_R$ space. These vevs are also chosen
in such a way that the  {\it modified horizontal survival hypothesis}\footnote{
The modified survival hypothesis states that, out of the known fermions, only
the top quark acquires tree level mass while the mass of the rest of these
fermions  is of radiative origin.} \cite{2,tuning} holds. The first steps of
the  SB chain arise from vevs of Higgs fields of the type $Z_3\phi(\ov
{n},1,n)$, where $n$ may be $15$ or $21$.

The SB chain is constrained by the requirement that the  evolution of the SM
coupling constants  from the unification scale to the scale $M_L$ of the last
step of  the chain, agrees with the experimental values~\cite{pdg} $sin^2
\theta_W   (M_L)=0.2315$, $\alpha_{EM}^{-1}(M_L)=127.9$,  $\alpha_3(M_L)=0.113$
and  $M_L\simeq10^2 GeVs$. For the renormalization group equations (rge), which
govern the evolution of the coupling constants, we adopt the {\it survival
hypothesis}\footnote{The survival hypothesis amounts to assume that at every
step, where a symmetry $G'$ is broken to $G''$ at the scale $M$, all fermions
whose mass is $G''$ invariant acquire mass of order $M$, with the possible
exception of the last step of the SB chain.} \cite{georgi} and the {\it
extended survival hypothesis}\footnote{For extended survival hypothesis it is
understood the assumption that the mass of all the Higgs scalars of the irreps
under $G'$ to which the scalars that acquire $G'$ breaking vevs belong, is of
order $M$. The rest of the scalars that complete an irrep under $G$ ($G'
\subset G$), have bigger masses  and are decoupled in the rge bellow M.}
\cite{aguila}.  When the symmetry is broken down to the SM group in $N$ steps
at the scales $M_k$, the coupling constants  satisfy, up to one loop, the rge
\be
 \alpha_i^{-1}(M_0) = f_i\, \alpha^{-1} - \sum^{N-1}_{k=0} b^k_i\,
\ln\left({M_{k+1}\over M_k}\right), 			\label{rge}
\ee
where $M_0=M_L$, $\alpha_i=g_i^2/4\pi$, $i =1$, 2, 3, and  $g_i$ are,
respectively, the gauge coupling  constants of the $U(1)_Y$, $SU(2)_L$ and
$SU(3)_c$ subgroups of the SM.  The factors $f_i$ are constants and define the
relation at the unification scale $M_N$ between $g$, the coupling constant of
$[SU(6)]^3\times Z_3$,  and $g_i$. The numerical  values of these factors, 
$f_1 = {14/3}$, $f_2 = 3$ and $f_3 = 1$~\cite{2,tuning}, arise from the
normalization conditions adopted for the generators of the corresponding gauge
group. In Eq. (\ref{rge})
 \be
 b^k_i = {1\over 4\pi}\left\{ {11\over 3}C^k_i({\rm vectors}) -
{2\over 3} C^k_i({\rm Weyl}\ {\rm fermions}) -
{1\over 6}C^k_i({\rm scalars})\right\},
\ee
where $C^k_i(\cdots)$ are the index of the representation to which the
$(\cdots)$ particles are assigned. For a complex scalar field the value of
$C^k_i({\rm scalars})$ should be doubled. The  relationships
 \be
\alpha_{EM}^{-1} \equiv \alpha_1^{-1} + \alpha_2^{-1}
\quad\mbox{ and } \quad  \tan^2\theta_W= {\alpha_1\over\alpha_2},
\ee
where $\theta_W$ is the weak mixing angle, hold  at all the energy scales and
from this expressions and Eq. (\ref{rge}) we have straightforwardly
\be
\alpha^{-1}_{EM}(M_0) - {23\over 3}\alpha_3^{-1}(M_0) =
\sum_{k=0}^{N-1}\left({23\over 3}b_3^k -b_1^k - b_2^k\right)
\ln\left({M_{k+1}\over M_k}\right) \label{awfer}
\ee
 and
\be
\sin^2\theta_W(M_0) = 3 \alpha_{EM}(M_0)\left[ \alpha_3^{-1}(M_0) +
\sum_{k=0}^{N-1}\left(b_3^k-{1\over 3}b_2^k\right)
\ln\left({M_{k+1}\over M_k}\right) \right].  \label{sinwfer}
\ee
Now, when  $N>1$ (the model with only two mass scales was excluded in Ref.
\cite{tuning,mass} by experimental data), the mass scales  and their hierarchy
are established by forcing the solutions of the rge to  agree with the
experimental data. In other words, the evolution of the coupling constants is
determined by the values of the intermediate scales, $M_k$, and the number of
them.

The type of topological defects depends then on the active symmetry at each
step. This is the subject of the next sections.


\section{Strings and domain walls.}
\eq

Cosmic strings, monopoles, and other topological structures can appear
during  phase transitions when a gauge group $\cal G$ is broken down
spontaneously to a subgroup $\cal H$ at certain energy scale $M_X$. The 
topological criterion for the existence of a string is the nontriviality
of the fundamental homotopy group of the vacua quotient manifold ${\cal M}
= {\cal G}/{\cal H}$, denoted by $\pi_1({\cal M}) (\not=0)$
\cite{preskill,tdvil,vacha}.
Similar to strings, the presence of other TD such as domain walls,
mono\-po\-les and textures is associated with the existence of a non trivial
homotopy group~\cite{preskill,tdvil,vacha} of the vacua manifold  $\cal M$. 
The $n$-$th$ homotopy group $\pi_n({\cal M})$ is the set of homotopically
equivalent classes of mappings from the $n$-sphere into the manifold $\cal M$. 
Domain walls exist when $\pi_0({\cal M})\neq 0$ while $\pi_2({\cal M})\neq 0$
means monopoles and $\pi_3({\cal M})\neq 0$ textures.
 In the case of a string, and at large distances
from it, the vacuum configuration of the scalar field, for
a connected and simply connected $\cal G$, is given by
\be
 \phi(\theta)=g(\theta)\phi_0, \qquad g(\theta) = e^{i\tau\theta} 
\ee
where $\tau$ is some generator of $\cal G$ not in the algebra of $\cal H$, 
$\theta$
is the azimuthal angle measured around the string, and $g(0)$ and $g(2\pi)$
belong to two disconnected pieces of $\cal H$. The presence of strings is then
signaled by the existence of noncontractible loops in the quotient space ${\cal
G}/{\cal H}$.
In what follows we  analize in a general
scheme the TD arising in the spontaneous SB of the model $[SU(6)]^3\times Z_3$.
We will see that some interesting topological aspects of Lie groups are
involved in the computation of TD.

Let ${\cal H}\rar {\cal G}\rar {\cal G}/{\cal H}$ be a generic Lie group
fibration which determines the spontaneous SB step at certain energy scale
$M_X$. The associated homotopy sequence to this fibration reads~\cite{mimu}
 \[\dots\rar\pi_n({\cal G})\rar\pi_n({\cal G/H})\rar\pi_{n-1}({\cal
 H})\rar\pi_{n-1}({\cal  G})\rar\dots
 \ .\]
For $n=1$ we obtain when ${\cal G}$ is connected ($\pi_0({\cal G})=0$) 
and simply
connected ($\pi_1({\cal G})=0)$
 \be 0\rar\pi_1({\cal G/H})\rar\pi_0({\cal H})\rar 0 \label{s3.3}\ee
or equivalently
 \be \pi_1({\cal G/H})\cong \pi_0({\cal H}), \ee
where ``$\cong$'' must be read as isomorphic. Therefore,
the only way to admit strings in whatever step of the SB is that the
corresponding residual group $\cal H$ must be not connected.

There are many possibilities for the chain of SB from $G = [SU(6)]^3\times Z_3$
down to  $SU(3)_c\otimes U(1)_{EM}$. Nevertheless,  most of them break $G$ down
to  connected groups.
An economical  SB scheme
 \be
 G\stackrel{M_R}{\Rar} G_1 \stackrel{M_H}{\Rar} G_{SM}
  \stackrel{M_L}{\Rar} G_r = SU(3)_c\otimes U(1)_{EM}, \label{eq1}
 \ee
where $G_{MS}$ is the SM group and $G_1 = SU(6)_L\otimes SU(4)_c\otimes
SU(2)_c\otimes SU(4)_R\otimes SU(2)_R\otimes U(1)$ was analyzed in
Ref.~\cite{thesis}. At each step however the residual symmetry group is
connected and therefore no strings are formed. The mass scales have
the hierarchy $M_R\sim 10^{11}\ GeVs\ > M_H\sim 10^{8}\ GeVs\ \gg M_L\sim 10^2\
GeVs\ $, which are in a good agreement with those obtained from the analysis 
of the generational see--saw mechanism in the model~\cite{seesaw}.  The last
pattern is implemented  by three sets of  Higgs fields in the irrep  $675$.
Their respective vevs are displayed in the second reference in~\cite{thesis}.

From here, the only way to break the symmetry down to a not connected group
$H$ is keeping the factor $Z_3$ intact in the first step. Nevertheless, as we
will show, there is only one way to do that and this SB scheme does not produce
strings.

As we  argued in the past section, we  choose the Higgs content to be of the
form  $Z_3\phi(\ov n,1,n)$, with  $ n=  15$ or $21$. Hence, if we require that
 \be
 G  = [SU(6)]^3\times Z_3 \rightarrow  H, \label{1.2}
 \ee
where $H$ contains the $Z_3$ symmetry unbroken, then, in order to respect this
symmetry, the tensor structure of $\la\phi\ra$ must be the same in the $L$, $R$
and $C$ spaces. Since ordinary color corresponds to $\alpha = 1,2,3$ in the
fundamental representation of $SU(6)_c$, the tensor indices of the terms in
$\la\phi\ra$ can be only $ 4, 5$ or $6$ in all the three spaces. On the other
hand $SU(2)_L$ should not be broken in the first step, that is $\la\phi\ra$
must take the direction of the singlets of $SU(2)_L,$
\[ \begin{array} {c c l}
  21 &:\qquad & \{1,4\}-\{2,3\},\ \ \{1,6\}-\{2,5\},\ \ \{3,6\}-\{4,5\},\\
& & \\
  15 &:\qquad  & [1,4]-[2,3], \ \ [1,6]-[2,5] \ \ [3,6]-[4,5],\\
& & [1,2] ,\ \ [3,4], \ \ [5,6],
\end{array} \]
since the $SU(6)_L$ (and $SU(6)_R$) indices are arranged according to the
following scheme:
\begin{equation}
\begin{array}{ c c c c}
  & \leftarrow& SU(3) & \rightarrow\\[1ex]
\uparrow & 1 & 3 & 5 \\
SU(2)  &   &   & \\
\downarrow & 2 & 4 & 6 \\
   \end{array}
\end{equation}
Therefore, the only Higgs field which maintains $Z_3$
unbroken is
$\phi_0\equiv\phi(675)=$ $Z_3\phi(\ov{15},1,15)$ with vevs in the direction
$[a,b],[\alpha,\beta],[A,B] = [5,6]$.

The symmetry breaking implemented by these vevs is (see appendix and Ref.
\cite{thesis})
\be
G \stackrel{\la\phi_0\ra}{\longrightarrow}
[SU(4)\otimes SU(2)]^3\otimes U(1)_{\Sigma}\times Z_3,
\ee
where $[G']^3\equiv G'_L\otimes G'_c\otimes G'_R$, $G' = SU(4)\otimes SU(2)$,
and $U(1)_{\Sigma}$ is generated by
\be
 T_{\Sigma} \equiv \frac{1}{\sqrt{3}} \left[ T \otimes 1\otimes 1 + 1\otimes
T \otimes 1 + 1\otimes 1\otimes T \right] \label{TSigma}
\ee
with
\be
 T=diag\{1,1,1,1,-2,-2\}/\sqrt{6}. \label{T}
\ee
Once we have fixed the first step of the SB pattern, the subsequent steps are
very constrained. In fact the more economical set of Higgs fields which
maintain this first step, and solve the rge in a proper way, uses six 675 to
break $H$ down to the SM in the following way (see the appendix)
\be
G\stackrel{M_G}{\Rar} H\stackrel{M_R}{\Rar}
I\stackrel{M_H}{\Rar}G_{SM}\stackrel{M_L}{\Rar} G_r \label{eq2}
\ee
where $M_X$ (X= G,R,H,L) represent the mass scale of the corresponding SB step
and
$ I = SU(4)_L\otimes SU(2)_L\otimes SU(4)_c \otimes SU(2)_c\otimes
Sp(4)_R\otimes  SU(2)_R\otimes U(1)_{\Sigma}$.
The hierarchy of the mass scales $M_G > M_R\geq M_H\gg M_L\sim 10^2GeVs$, is
enough to cope with the experimental results. Precise values of $M_G$, $M_R$
and $M_H$ are obtained from the rge (\ref{awfer} and \ref{sinwfer}).  Figures 1
and 2 show the corresponding ones of $M_G$ and $M_R$ as a function of  the
values taken  for $M_H$, when we use five or six 675 to make the breaking down
to the SM. Also we plotted the identity function for $M_H$, looking for the
limit point $M_R=M_H$  where we have just an intermediate scale between $M_G$
and $M_L$. As it can be noted, the intermediate scale is necessary to secure
that $M_G\leq 10^{19}GeVs$, in the more economical case, but not in the other
one.

\begin{figure}[hc]
\epsfxsize = 260pt
\centerline{\epsfbox{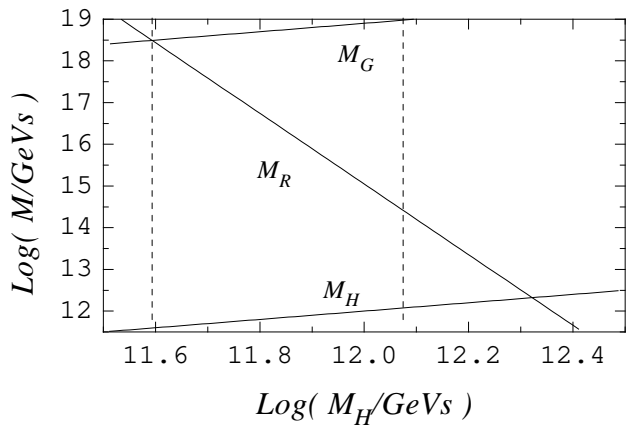}}
\vspace{-2em}
\quote{\caption{Evolution of the mass scales $M_G$, $M_R$ as a function of
$M_H$ when five 675 are used to break $G$ down to the SM. Notice that the point
$M_R = M_H$ is out of the window $M_H\leq M_R<M_G\leq 10^{19}\ GeVs$.}}
\end{figure}
\begin{figure}
\epsfxsize = 260pt
\centerline{\epsfbox{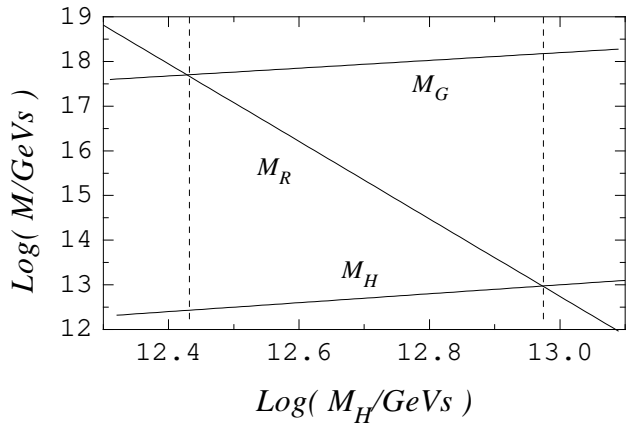}}
\quote{\caption{Development of the mass scales when we use six 675
 for the SB. Now the point $M_R = M_H$ is consistent
 with $M_G\leq{10^{19}\ GeVs}$. }}
\end{figure}

Now, $G$ in the present case is a not connected group due the factor
$Z_3$, then the sequence (\ref{s3.3}) is not longer valid. Instead we have
the complete homotopy sequence for $n=1$
 \be \pi_1(G)\rar\pi_1(G/H)\rar\pi_0(H)\rar\pi_0(G). \ee
In other words, for the case considered above, we have
 \[0\rar\pi_1(G/H)\rar Z_3\rar  Z_3.\]
Here we have used the usual relation $\pi_i(X \times Y) = \pi_i(X) \oplus
 \pi_i(Y)$ for any $X,Y$ topological spaces and the following
theorem~\cite{mimu}:

Let $Q$ be a discrete finite group of $N$ elements, then $\pi_n(Q)=0$ for
all $n\neq 0$ and $\pi_0(Q)\cong Z_N$. Here $Z_N$ is the finite cyclic
group of order $N$.

In order to compute $\pi_1(G/H)$ we would like first to explore the
topology of the quotient manifold ${\cal M} = G/H.$ It is easy to see that
globally it looks like the product of two complex Grassmannians
$G_4({\com}^6)$ and the quotient manifold $G_4({\com}^6)/U(1)_{\Sigma}$.
Thus the quotient ${\cal M}= G/H$ looks topologically like
\be {\cal M} = G/H = G_4^L(\com^6)\times G_4^c(\com^6)\times
G_4^R(\com^6)/U(1)_\Sigma. \label{gras}\ee
Here $G_4({\com}^6)$ is represented as the homogeneous space
 \be G_4({\com}^6) = {SU(6)\over SU(4)\times SU(2)}. \ee
The general case is
 \be G_k({\com}^n)\equiv  {SU(n)\over SU(n-k)\times SU(k)}. \ee

It is very well known ~\cite{theorem} that the complex Grassmannian is a
compact, connected and simply connected homogeneous space and the same
occurs with a complex Grassmannian divided by any compact, connected and
simply connected space (in our case it is $U(1)_{\Sigma}$ which is
topologically the circle). Thus we conclude that
  \be \pi_1({\cal M}) = 0.\ee
and therefore there are not strings at the scale $M_G$. From the structure
of the SB chain (\ref{eq2}) the presence of strings might only occurs at
the first SB step. 
We saw that it does not occurs  and therefore there is no way to produce stable
strings in the model. Nevertheless, metastable strings could be formed during 
a intermediate phase transition if the sequence
\be
  {\cal G}'\subset{\cal G} \rightarrow {\cal G}'' \otimes U(1) \otimes U(1)' 
  \rar {\cal G''} \otimes U(1) \label{a}
\ee
could be implemented in the model~\cite{tdvil,16}. In this case the strings
connecting monopole-antimonopole pairs are associated to the broken $U(1)'$
symmetry and therefore they might decay by nucleation of pairs~\cite{vil3}
giving a possible contribution to the HECR events.

Effectively, a sequence of the type (\ref{a}) may be implemented in the
model as it follows from the analysis done in reference~\cite{tuning,thesis}
\be
    G \stackrel{M_R}{\Rar} G_1 \stackrel{M_H}{\Rar}
    G_{SM} \otimes U(1)' \stackrel{M_S}{\Rar}G_{SM}
    \stackrel{M_L}{\Rar} G_r.  \label{b}
\ee

In the sequence (\ref{b})  the mass scales have the hierarchy $M_R\sim 10^{12}$
GeV$>M_H\sim 10^8$ GeV$> M_S> M_L$. The scale  $M_S$ corresponds to the phase
transition at which metastable strings are formed. $M_S$ is not fixed by the
rge since the scalar fields which breaks the $U(1)'$ symmetry are singlets of
the SM group.

On the other hand, the presence of domain walls formed during the phase
transitions of a universe with the gauge symmetry $\cal G$ is a cosmological
problem which can not be solved, then, a realistic model should not
predict them in order to be considered as admissible. In our case, we have
to compute the homotopy group which determines their presence. That means
 to compute $\pi_0({\cal M})$ for all the steps of the SB chains.

A second consequence from the topological structure of ${\cal M}$ Eq.
(3.12) is that
   \be \pi_0({\cal M}) = 0.\ee
Then there are no domain walls at the $M_R$ scale.

For the second step $H\rar I$, the only broken symmetry is $SU(4)_R$ whose
residual symmetry is $Sp(4)_R$, so, we compute easily $\pi_0(H/I)$ to be
trivial, and conclude that again there are no domain walls formed in this step.
This is because $SU(4)$ and $Sp(4)$ are connected, simply connected and compact
Lie groups.  In the next step, $I\rar G_{SM}$, the calculation gives
$\pi_0(I/G_{SM})=0$, and finally the breaking of the SM symmetry does not
produce domain walls. Hence, they are absent in the chain (\ref{eq2}).

The same analysis for the minimal chain (\ref{eq1}) gives
 \be \pi_0(G/G_1)=\pi_0(G_1/G_{SM})=0,\ee
and in a similar way from sequence (\ref{b}), $\pi_0 (G_1/G_{SM}\times U(1)') =
\pi_0 (G_{SM} \times U(1)'/G_{SM}) = 0$, because all involved groups are
compact. So, we can conclude that the model is free of domain walls. Notice
that in all computations of the above homotopy groups  we have used the
standard techniques of semistable homotopy groups of Ref. ~\cite{mimu}.

\section{Monopoles in $[SU(6)]^3\times Z_3$.}
\eq

Monopoles are the more common kind of topological defects in GUM's. As we
mentioned above, they are associated with the existence of a non trivial
homotopy group: $\pi_2({\cal M})\neq 0$ \cite{preskill,tdvil,vacha}. Then, the
presence of $U(1)$ factors in the residual symmetry after the breaking is
sufficient to show the formation at that step of monopoles in the model, trough
the Kibble mechanism~\cite{kibble}.

By construction, all the GUM's produce monopoles, even when they break the
unified symmetry to the SM group in a simple step, because the SM group
contains a $U(1)_Y$ factor associated to the hypercharge. Then if the
conventional philosophy of the GUT's is correct, monopoles must exist unless
there is some unknown mechanism which kills them. Of course an obvious way to
relax this problem is including a $U(1)_Y$ factor at the level of the GUT, but
in this case the model has not a simple gauge coupling in a natural way.  This
is a motivation to consider monopoles as a good candidates to be sources of the
HECR~\cite{sigl1,sigl2,chi,kephart,he4}. Monopolonium~\cite{hill} is a possible
bound state formed from a monopole-antimonopole pair, which spirals in and
finally collapses. This is a very slow process which fails in solving the
monopole overabundance of the early universe. Nevertheless, if by some
mechanism the universe is never monopole dominated, or also, if they exist at
the level of abundance compatible with known experimental~\cite{ahlen} and
phenomenological upper bo\-unds~\cite{parker}, then the late annihilation of
monopoles could be precisely the mechanism that we need from the point of view
of generating HECR, which must be produced only in the contemporary cosmic
epoch. Besides the associated symmetry, the only clear difference between 
monopoles produced by a specific GUM with respect to another one is the mass
scale at which they appear, because it depends strongly in the SB pattern. In
most of the models which achieve the simple unification, the scale is of the
order of $10^{16}\ GeVs$. If this is the case, the production of HECR based on 
monopole acceleration must be ruled out, because of the upper bound for the
monopole mass $M\leq 10^{10\pm 1}\ GeVs$~\cite{kephart} obtained from  the
phenomenological and experimental bounds to their flux~\cite{parker,ahlen}.

In the case we are considering here, the minimal pattern of the SB given by
(\ref{eq1}) produces monopoles at all the scales, Then the more massive ones
should appear around $M_R\sim 10^{11}\ GeVs$. Also monopoles are formed at the
mass scales $M_H\sim 10^8\ GeVs$ and $M_L\sim 10^2\ GeVs$. Now, monopole
annihilation could not give  an important contribution to the HECR in this
model, but instead the acceleration mechanism could be accepted, even if the
mass of the $10^{11}\ GeVs$ monopoles get in conflict with the experimental
data, because, in such a case one can invoke inflation below or at that scale.
In this case, the massive monopoles should have no more relevance for the HECR,
but the lighter ones should produce HECR by the acceleration mechanism. In a
similar way, in the pattern given by (\ref{b}) monopoles appear at the same
scales, but now two classes of them are produced at the scale $M_H$. Those
associated to the broken symmetry $U(1)'$ become attached to strings as we
argued in the above section. The other monopoles should be free and they could
be accelerated to relativistic energies just like those produced in the pattern
(\ref{eq1}).

On the other hand, the exotic chain (\ref{eq2}), has a higher unification
scale ($M_G\sim 10^{18}\ GeVs$) than the more usual patterns, and also
than the minimal chain in (\ref{eq1}). For the model with this SB scheme
    \be \pi_2(G/H)\cong Z. \ee
Then we have supermassive monopoles at the same scale $M_G$. Now, the
monopoles produced by this phase transition  may give an important contribution
to the HECR via the monopole-antimonopole annihilation mechanism.  As we
discuss before, this kind of monopoles could overclose the early universe, and
therefore they constitute a serious  cosmological problem. even so, as Sigl
{\it et. al.} have argued~\cite{sigl1,sigl2,chi}, from numerical simulations,
it looks that it is  possible (with proper assumptions) that they could give a
substantial contribution to the HECR flux. Notice that the chain (\ref{eq2}) is
the only known in this model with a unification scale as high as the one of a
typical GUM. Hence, this SB pattern should be ruled out if the monopolonium
annihilation mechanism is not confirmed by future observations like those
planed in the ``Pierre Auger Project'', which expects to collect six thousand
events per year above $10^{19} \ GeVs$ in a 6,000 $Km^2$ detector~\cite{auger}.

At the second step of the chain (\ref{eq2}) the $U(1)_\Sigma$ factor stays
intact and not extra $U(1)$ factors arise, then $\pi_2(H/I) = 0$,  as
the homotopy computation shows. Hence, no monopoles are formed at that
step, The nexts steps produce residual $U(1)$ symmetries. Then, as an easy
calculations indicates, we have monopoles at the scales $M_H$ and $M_L$, for
this SB scheme.

Finally, we have to mention that local textures are formed from the phase
transitions of the model in both schemes [Eq (\ref{eq1}) and (\ref{eq2})].
We compute $\pi_3({\cal M})$ for all the steps and conclude that local
textures appear only at the scales $M_H$ and $M_L$ in both SB patterns.
However, their possible relation with the sources of the HECR is not clear
until now.

\vskip 1truecm

\section{Concluding remarks.}

In this paper we have started the study of the TD arising in the model proposed
in Refs.~\cite{oax,2,tuning}. 
We have explored the possibility of breaking
$[SU(6)]^3 \times Z_3$ down to the SM group through a chain that would give
rise to  the formation of stable strings, however  our conclusion is thta it is
not possible. 
Nevertheless, in certain patterns [Eqs. (\ref{a}) and (\ref{b})]
the model does admit metastable strings which connect monopole-antimonopole
pairs. Basically, we have established that this vacuum configuration could
appear at an scale below $M_H\sim 10^8$ GeV. The decaying processes which
involve these kind of hybrid defects could give a relevant contribution to the
HECR flux.  Also, the homotopy analysis of the minimal scheme (\ref{eq1}) of
the SB  chain and of the exotic one (\ref{eq2}) shows that domain walls are
absent, which is in  very good agreement with cosmological restrictions. The
same is valid in the case of the pattern (\ref{b}) which produces metastable
strings. Hence, this result could give us an indirect test of the viability of
the model, besides its special properties mentioned along the paper. Also, the
model contains local textures, but they are not of our interest  now, because
there is not a clear mechanism that involves them as sources of the HECR.

Other TD relevant  for the HECR problem produced in the model  are monopoles,
whose energy scale depends strongly on the specific pattern chosen for the SB.
We have analyzed here two schemes. While the minimal chain produces monopoles
at all the relevant energy scales  $M_R\sim 10^{11}\ GeVs\ > M_H\sim 10^{8}\
GeVs\ \gg M_L\sim 10^2\ GeVs$, in the new chain (\ref{eq2}) they are formed at
higher energies, $M_G\sim 10^{18}\ GeVs$ and $M_H\sim 10^{12.5}\ GeVs$, and of
course also at the SM scale $M_L$. As a matter of fact, if the annihilation
mechanism is correct and it is confirmed by future data, then the model does
not has  any advantage over  other GUM's because  most of them predict the
formation of monopoles at or above $10^{16}\ GeVs$, and then only a very
precise analysis could rule out any of these models. On the other hand, the
presence of monopoles in the more natural SB scheme given by (\ref{eq1}) at not
so high energies,  could actually give to the model a real advantage over other
GUM's in the case that the HECR should be produced by the acceleration 
mechanism~\cite{kephart}.  This is by the moment, an open problem to be
analyzed in the future.   Another open problem is the relation of the results
obtained in this paper and the recent measurments of the extragalactic diffuse
gamma ray background~\cite{edgrb} which also seems to be difficult of
explaining otherwise in terms of emissions from astrophysical objets. This
problem has been recently studied in the context of supersymmetric
GUT's~\cite{battha2}, where strings formed at one scale below $10^{14}$ GeV
could simultaneously be source of high energy particles (with masses of the
order of the breaking scale) and low energy higgs particles (with masses as low
as 1 TeV). In these scenarios, while the high energy particles may contribute
to the HECR flux, the low energy higgs` decay can account for the extragalactic
diffuse gamma ray background.  However, in a non supersymmetric GUT, the
possibility of implement this mechanism is still an open question.   In
supersymmetric theories the basic ingredient are the  flat directions of the
scalar potential, which is absent in the non supersymmetric theories. However,
the formation of hybrid defects at low energy scales, as in the case of the
chain (\ref{b}) where scalars could appear at low energy may give such
contributions, but more analysis is required to establish this possibility.

Finally the TD discussed along the paper for the chains (\ref{eq1}) and
(\ref{eq2}) are not metastable but topologically stable. This is due to the
fact that they do not satisfy the Preskill-Vilenkin criterion~\cite{16}.

\vskip 1truecm

\section*{Acknowledgments}

This work was partially supported by CONACyT in M\'exico. One of us (A.Z)
acknowledges the hospitality of Prof. J. Bernabeu and of the Theory Group
at the University of Valencia as well as the financial, support during the
1995-1996 sabbatical leave, of Direcci\'on General de Investigaci\'on
Cient\'{\i}fica y T\'ecnica (DGICYT) of the Ministry of Education and
Science of Spain. We acknowledge W. A. Ponce for his helpful discussions
and comments, G. Moreno for much help and advice in the computations
of the homotopy of Lie groups, and M. Boratav for drawing our attention to
Ref.~\cite{kephart}.

\vskip 1truecm

\appendix

\section*{Appendix A}
\eq

In this appendix we give some details about the SB
implemented in section 3. First, since
$\la\phi_0\ra$ is $Z_3$--invariant, it induces
the breaking
        \[ G\rightarrow G'\times Z_3  \]
where $G'$ is a $Z_3$--invariant subgroup of $[SU(6)]^3$.

Next, consider the group $SU(4)\otimes SU(2) \subset SU(6)$, where
 $SU(4)$ ($SU(2)$) acts only on the indices $1,2,3,4$ ($5,6$) of the
fundamental representation of $SU(6)$.
$SU(4)$ ($SU(2)$) is the
largest simple group contained in $SU(6)$ which acts only on the subspace
with tensor indices $1,2,3,4$ ($5,6$). For this decomposition
of the space, the branching rules read
\beas
SU(6) &\rightarrow& SU(4)\otimes SU(2)\\
6 &\rightarrow& (4,1) + (1,2)\\
15 &\rightarrow& (1,1) + (6,1) + (4,2) \\
21 &\rightarrow& (10,1) + (1,3) + (4,2)\\
35  &\rightarrow& (1,1) + (15,1) + (1,3) + (4,2) + (\bar 4, 2),
\eeas
which shows that there is only one component in the irrep 15 of $SU(6)$  which
is  singlet of $SU(4)\otimes SU(2)$  and that  it corresponds precisely to
the direction  $[5,6]$, which is the direction in the $L$, $R$, and $c$ spaces
where we  have demanded $\langle\phi_0\rangle \neq 0$. On the other hand the
maximal subgroup of $SU(6)$ which contains $SU(4)\otimes SU(2)$ is
$SU(4)\otimes SU(2)\otimes U(1)$ where $U(1)$ is generated by the matrix
displayed in eq. (\ref{T}). Obviously these $U(1)$ factors in $G$ are  broken
by $\langle\phi_0\rangle$, but the sum of their generators,  $T_{\Sigma}$,
given in eq. (\ref{TSigma}) satisfies
\[
T_{\Sigma} \langle\phi_0\rangle = 0
\]
and therefore $U(1)_{\Sigma}$ generated by $T_{\Sigma}$ remains unbroken.
Therefore, the SB induced by
$\la\phi_0\ra$ is
\[
  G\rightarrow [SU(4)\otimes SU(2)]^3\otimes U(1)_{\Sigma}\times Z_3.
\]

It is important to note, that now in this chain, the ordinary color group
$SU(3)_c\subset SU(4)_c$, and the the left group $SU(4)_L\otimes SU(2)_L$
decomposes in $SU(2)_L$ $\otimes SU(2)_{HL}$, where the last $SU(2)_L$ is the
standard electroweak group and the horizontal group $SU(2)_{HL}$ acts only
in the space of the the first two families, as it follows from the
branching rules
\beas
SU(4)_L\otimes SU(2)_L &\rightarrow& SU(2)_L\otimes SU(2)_{HL}\\
(4,1) &\rightarrow& (2,2)\\
(1,2) &\rightarrow& (2,1).
\eeas
The subsequent steps of the SB were chosen in order to get the highest
contribution  from the Higgs sector to the rge. The vevs of the Higgs fields
used for next steps of the  SB are
\[
\la\phi_{1[a,b]}^{[A,B]}\ra =\la{\phi'}_{1[a,b]}^{[A,B]}\ra =
\la{\phi}_{2[a,b]}^{[A,B]}\ra = 0,
\]
and
\[\ba{ll}
\la\phi^{[a,b]}_{1[\alpha,\beta]}\ra =
 \la\phi_{1[A,B]}^{[\alpha,\beta]}\ra =
M_H &\mbox{ for }[a,b], [A,B] = [1,2]=[4,5]=-[3,6], \\ [6pt]
&\mbox{ and }[\alpha,\beta]=[4,5],
\ea\]
\[\ba{ll}
\la{\phi'}^{[a,b]}_{1[\alpha,\beta]}\ra =
 \la{\phi'}_{1[A,B]}^{[\alpha,\beta]}\ra =
M_H & \mbox{ for }[a,b], [A,B]=-[1,2]=[4,5]=-[3,6],\\ [6pt]
&\mbox{ and } [\alpha,\beta]=[4,5], \\ [6pt]
\la{\phi}^{[a,b]}_{2[\alpha,\beta]}\ra =
 \la{\phi}_{2[A,B]}^{[\alpha,\beta]}\ra =
M_H & \mbox{ for }[a,b], [A,B]=-[2,5]=[1,6]=[3,4],\\ [6pt]
&\mbox{ and } [\alpha,\beta]=[4,5],
\ea\]
and finally
\[\ba{ll}
\la{\phi'}^{[a,b]}_{2[\alpha,\beta]}\ra =
M_H & \mbox{ for }[a,b]=-[2,5]=[1,6]=-[3,4],
\mbox{ and } [\alpha,\beta]=[4,5],\\ [6pt]
\la{\phi'}_{2[A,B]}^{[\alpha,\beta]}\ra =
M_H & \mbox{ for }[A,B]=[2,6] \mbox{ and } [\alpha,\beta]=[4,6],\\ [6pt]
\la{\phi'}_{2[a,b]}^{[A,B]}\ra = M_R & \mbox{ for }[A,B]=[2,3]=-[1,4]
 \mbox{ and } [a,b]=[5,6].
\ea\]
They, together with $\la\phi_4\ra$, are sufficient to implement the SB
(\ref{eq2}).  We
must however add one or more Higgs fields, in order to solve consistently the
rge. Those fields are chosen to take vevs along the directions of the $(4,2) +
(6,1)$ irrep of $SU(4)_{L,R}\otimes SU(2)_{L,R}$, and the  $(4,2)$ of
$SU(4)_c\otimes SU(2)_c$ in such a way that $\la \phi^{[A,B]}_{[a,b]}\ra = 0$.


\end{document}